\documentclass[%
 reprint,
 groupedaddress,
 showpacs,
 amsmath,amssymb,
 aps,
 prb
]{revtex4-1}

\usepackage[dvipdfmx]{graphicx, color}
\usepackage[dvipdfmx]{hyperref}
\usepackage{bm}

\graphicspath{{./images/}}
\DeclareMathOperator{\Tr}{Tr}
\DeclareMathOperator{\sgn}{sgn}
\begin{document}

\title{Superconductivity in magnetic multipole states}

\author{Shuntaro Sumita}
\email[]{s.sumita@scphys.kyoto-u.ac.jp}
\affiliation{%
 Department of Physics, Graduate School of Science, Kyoto University, Kyoto 606-8502, Japan
}%

\author{Youichi Yanase}
\affiliation{%
 Department of Physics, Graduate School of Science, Kyoto University, Kyoto 606-8502, Japan
}%


\date{\today}

\begin{abstract}
Stimulated by recent studies of superconductivity and magnetism with local and global broken inversion symmetry, we investigate the superconductivity in magnetic multipole states in locally noncentrosymmetric metals.
We consider a one-dimensional zigzag chain with sublattice-dependent antisymmetric spin-orbit coupling and suppose three magnetic multipole orders: monopole order, dipole order, and quadrupole order.
It is demonstrated that the Bardeen-Cooper-Schrieffer state, the pair-density wave (PDW) state, and the Fulde-Ferrell-Larkin-Ovchinnikov (FFLO) state are stabilized by these multipole orders, respectively.
We show that the PDW state is a topological superconducting state specified by the nontrivial $\mathbb{Z}_2$ number and winding number.
The origin of the FFLO state without macroscopic magnetic moment is attributed to the asymmetric band structure induced by the magnetic quadrupole order and spin-orbit coupling.
\end{abstract}


\maketitle


\section{Introduction}
Emergent phenomena in electron systems lacking inversion symmetry have received a lot of attention in recent condensed matter physics~\cite{NoncentroSC, Nagaosa2013}.
In such noncentrosymmetric systems, antisymmetric spin-orbit coupling (ASOC) entangles various internal degrees of freedom: for instance, spin, orbital, sublattice, and multipole.
Recent studies uncovered exotic superconducting~\cite{Yoshida2012, Yoshida2013, Yoshida2015, Watanabe2015} and multipole phases~\cite{Spaldin, Yanase2014, Hitomi2014, Hayami2014-02, Hayami2014-08, Hayami2015, LiangFu2015, Matteo} induced by the sublattice-dependent ASOC in locally noncentrosymmetric systems.
In this paper, we clarify nontrivial interplay between the superconductivity and the multipole order by investigating the superconductivity in the magnetic multipole states. 

Even-parity multipole order has been intensively researched mainly in the field of heavy-fermion systems.
For instance, the electric quadrupole and magnetic octupole order have been identified in various materials~\cite{Kuramoto_review}.
Furthermore, the electric hexadecapole moment~\cite{Haule2009,Kusunose2011} and magnetic dotriacontapole moment~\cite{HIkeda2012} have been proposed as plausible candidates for the hidden order parameter in the heavy-fermion superconductor (SC) URu$_2$Si$_2$.

On the other hand, recent theories~\cite{Spaldin, Yanase2014, Hitomi2014, Hayami2014-02, Hayami2014-08, Hayami2015, LiangFu2015} pointed out the odd-parity multipole order which may occur in the locally noncentrosymmetric systems as a result of the antiferro alignment of the even-parity multipole in the unit cell.
For instance, the ``antiferromagnetic moment'' in the unit cell induces a magnetic quadrupole moment~\cite{Yanase2014, Hayami2015}, and the antiferro stacking of the local electric quadrupole moment in bilayer Rashba systems is regarded as an electric octupole order~\cite{Hitomi2014}.
As a consequence of the spontaneous global inversion symmetry breaking, intriguing magnetoelectric responses occur in the ferroic odd-parity multipole states~\cite{Spaldin, Yanase2014, Hitomi2014, Hayami2014-02, Hayami2014-08, Hayami2015, LiangFu2015, Matteo}.
Recent experiments detected a signature of the odd-parity multipoles in Sr$_2$IrO$_4$~\cite{zhao2016, Matteo}.
Inspired by these works, we study exotic superconductivity induced by the odd-parity multipoles and even-parity multipoles.

Intensive theoretical studies in these years have shown that noncentrosymmetric SCs are platforms of various nonuniform superconducting states~\cite{NoncentroSC}.
In the globally noncentrosymmetric systems an infinitesimal magnetic field stabilizes a Fulde-Ferrell-Larkin-Ovchinnikov (FFLO) state~\cite{FF, LO}, which is called the helical superconducting state~\cite{Edelstein, Dimitrova, Samokhin}.
Agterberg and Kaur discussed the stability of the magnetic-field-induced FFLO (helical) state in Rashba SCs~\cite{Agterberg}.
However, it has been shown that the FFLO order parameter is hidden in vortex states~\cite{Matsunaga, Hiasa}.

In the locally noncentrosymmetric systems, the pair-density-wave (PDW) state~\cite{Yoshida2012} or the complex stripe state~\cite{Yoshida2013} may be stabilized, depending on the direction of magnetic field.
These states are not hidden in the vortex states, but a magnetic field higher than the Pauli-Chandrasekhar-Clogston limit is required. 
Reference~\onlinecite{Yoshida2012} has shown that the PDW state is stable in multilayer SCs having ``weak interlayer coupling'' and ``moderate spin-orbit coupling'' when the paramagnetic depairing effect is dominant.
Then, the phase of the superconducting order parameter modulates layer by layer.
Therefore, the PDW state is an odd-parity superconducting state although the spin-singlet Cooper pairs lead to the condensation.
Since the odd-parity SC is a platform of topological superconducting phases~\cite{Sato2010}, topologically nontrivial properties of the PDW state are implied.
Indeed, the PDW state in 2D multilayer systems has been identified as being a crystal-symmetry-protected topological superconducting state~\cite{Yoshida2015, Watanabe2015}.

\begin{figure}[h]
 \centering
 \includegraphics[width=90mm, clip]{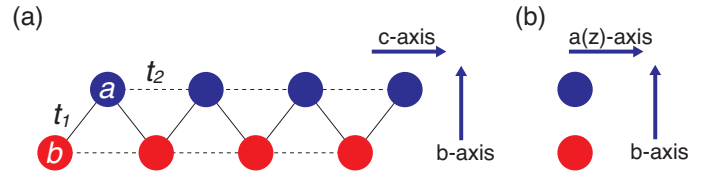}
 \caption{Crystal structure of the 1D zigzag chain. (a) Projection along the \textit{a} axis. (b) Projection along the \textit{c} axis. Blue and red circles represent the $a$ and $b$ sublattices, respectively. The hopping integrals are shown by $t_1$ and $t_2$.}
 \label{fig:1d-zigzag}
\end{figure}

The previous theories introduced above discussed the superconducting state in the magnetic field.
In this paper, we investigate the superconductivity caused by the cooperation of various magnetic multipoles and sublattice-dependent ASOC.
Since this is an early theoretical study for those systems, we treat a one-dimensional (1D) zigzag chain (Fig.~\ref{fig:1d-zigzag}) as a minimal model.
Indeed, the zigzag chain is a simple crystal structure lacking the local inversion symmetry.
Although the superconducting long-range order does not occur in strictly 1D systems because of divergent fluctuations~\cite{Mermin-Wagner}, we investigate superconducting states with the use of the mean-field (MF) theory (Sec.~\ref{sec:mf_theory}) by allowing the long-range order.
This treatment is appropriate for our purpose to pave a way to realize exotic superconductivity in a broad range of systems.
Indeed, our results are justified in quasi-1D coupled zigzag chains, and some of the results will give new insight on more complicated three-dimensional (3D) systems with broken local inversion symmetry.

\begin{figure}[htbp]
 \centering
 \includegraphics[scale=0.45,clip]{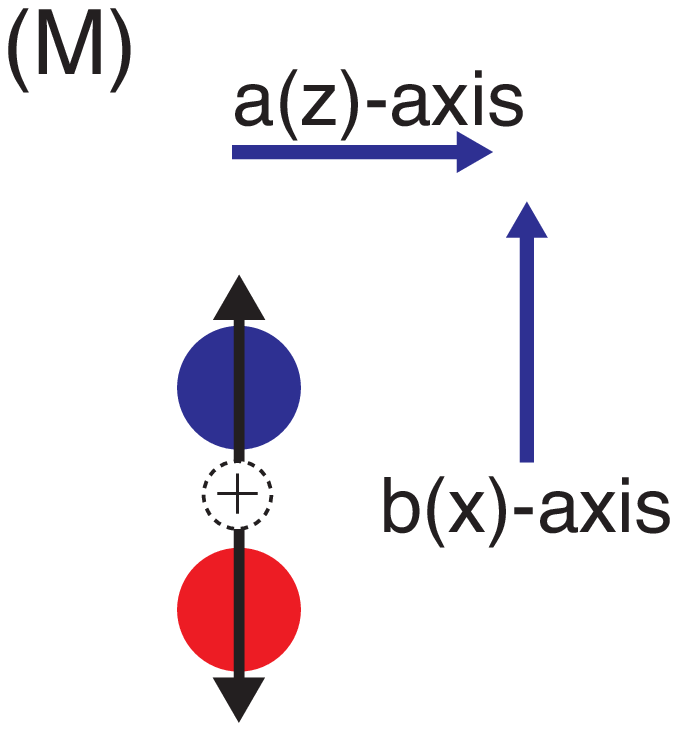}
 \includegraphics[scale=0.45,clip]{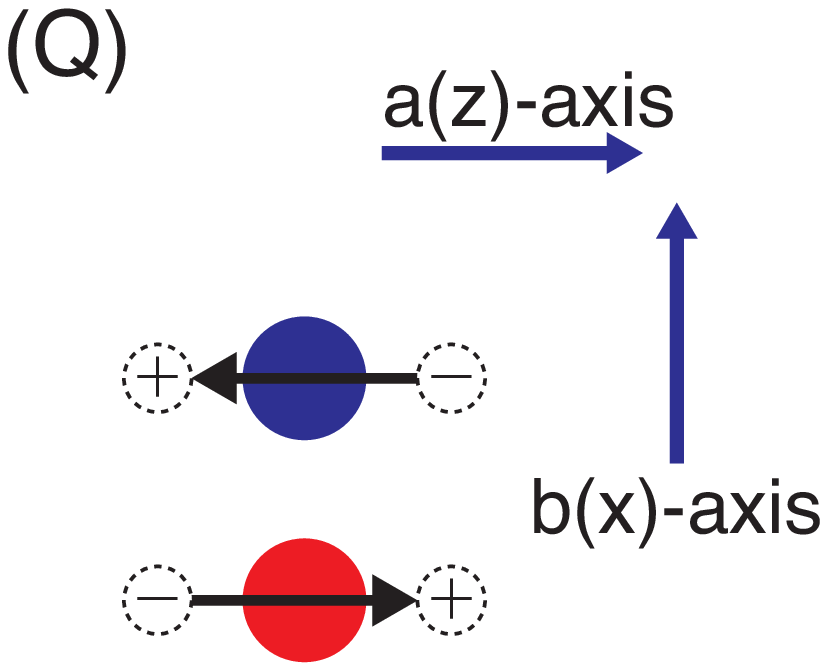}\\
 \includegraphics[scale=0.45,clip]{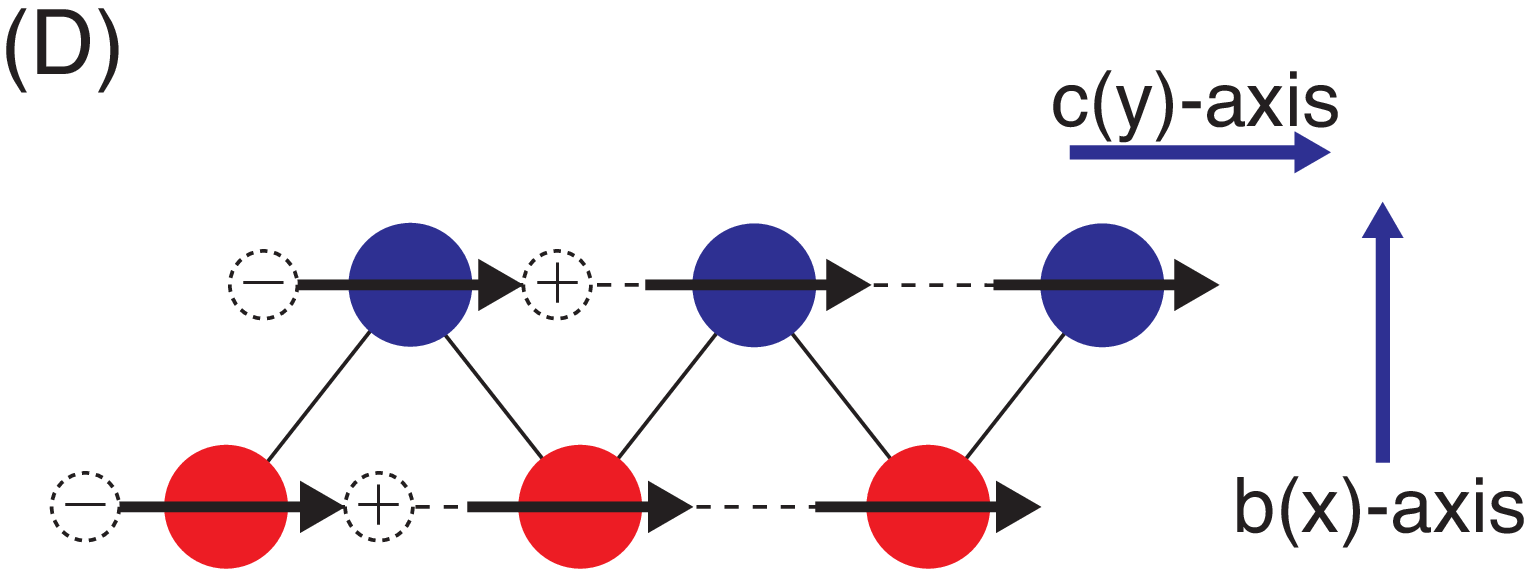}
 \caption{Magnetic structure in the magnetic (M) monopole state, (D) dipole state, and (Q) quadrupole state. Projection along the \textit{c} axis is shown in (M) and (Q), meanwhile along the \textit{a} axis is shown in (D). Black arrows show the ferromagnetic or ``antiferromagnetic'' moments in the unit cell.}
 \label{fig:magnetic_structure}
\end{figure}

We show that the PDW state is stabilized in the magnetic dipole state [Fig.~\ref{fig:magnetic_structure}, panel (D)] as in the multilayer systems, while the conventional Bardeen-Cooper-Schrieffer (BCS) state is robust in the magnetic monopole state [Fig.~\ref{fig:magnetic_structure}, panel (M)].
Topologically distinct properties of the PDW state are specified by the $\mathbb{Z}_2$ and $\mathbb{Z}$ topological invariants.
The Majorana end state is associated with nontrivial topological invariants.
In this sense, the odd-parity PDW state is regarded as a realization of the Kitaev superconducting wire~\cite{Kitaev2001} without $p$-wave Cooper pairs.
Ferromagnetic heavy fermion SCs, UGe$_2$~\cite{Saxena2000}, URhGe~\cite{Aoki2001}, and UCoGe~\cite{Huy2007} have crystal structure consisting of zigzag chains, and they are likely to show the odd-parity superconductivity.
Therefore, these compounds are candidates for the PDW state.

Furthermore, we show that the FFLO state is stable in the magnetic quadrupole state [Fig.~\ref{fig:magnetic_structure}, panel (Q)] without macroscopic magnetization.
The magnetic quadrupole order occurs in several materials.
For example, 1-2-10 compounds such as CeRu$_2$Al$_{10}$ show magnetic quadrupole order in zigzag chains~\cite{Khalyavin2010, Tanida2011, Muro2011, Kato2011}.
Because any external magnetic field is not required for the FFLO state, the orbital effect harmful for the FFLO state~\cite{Matsunaga, Hiasa} is completely eliminated.
Thus, the magnetic quadrupole state is a good platform realizing the FFLO state which has been searched for more than 50 years~\cite{Bianchi2003, Kenzelmann2008, Uji2006, Lortz2007}.

This paper is constructed as follows.
In Sec.~\ref{sec:model}, we introduce a model for conduction electrons affected by a sublattice-dependent ASOC, magnetic multipole order, and an $s$-wave attractive interaction.
Then, we analyze the model with the use of the MF theory in Sec.~\ref{sec:mf_theory}.
We illustrate the monopole, dipole, and quadrupole order in Sec.~\ref{sec:non-interacting}.
The symmetry and degeneracy of the band structure are elucidated by clarifying the symmetry protection.
In Sec.~\ref{sec:order_parameter}, we introduce the order parameter of superconducting states discussed in this paper.
We show that the BCS state is robust against the magnetic monopole order in Sec.~\ref{sec:monopole}.
On the other hand, the dipole order stabilizes the PDW state as shown in Sec.~\ref{sec:dipoleA}.
The PDW state is identified to be a topological superconducting state in a certain parameter regime (Sec.~\ref{sec:topological}).
Section~\ref{sec:quadrupole} gives the result for the FFLO state induced by the quadrupole order despite the absence of the external magnetic field.
It is shown that the center-of-mass momentum of Cooper pairs arises from the asymmetric band structure.
Finally, a brief summary and discussion are given in Sec.~\ref{sec:summary}.

\section{Model and Formulation}

\subsection{Model}
\label{sec:model}
First, we introduce a model describing superconductivity coexisting with magnetic order in a 1D zigzag chain,
\begin{align}
 {\cal H} =& \sum_{k, s} [ \varepsilon(k) a_{k s}^\dagger b_{k s} + \text{h.c.} ] \notag \\
 & + \sum_{k, s} [ \varepsilon'(k) - \mu ] [ a_{k s}^\dagger a_{k s} + b_{k s}^\dagger b_{k s} ] \notag \\
 & + \alpha \sum_{k, s, s'} \bm{g}(k) \cdot \bm{\hat{\sigma}}_{s s'} [ a_{k s}^\dagger a_{k s'} - b_{k s}^\dagger b_{k s'} ] \notag \\
 & - \sum_{k, s, s'} [ \bm{h}_a \cdot \bm{\hat{\sigma}}_{s s'} a_{k s}^\dagger a_{k s'} + \bm{h}_b \cdot \bm{\hat{\sigma}}_{s s'} b_{k s}^\dagger b_{k s'} ] \notag \\
 & + \frac{1}{N} \sum_{k, k', q} V_a(k, k') a_{k + \frac{q}{2} \uparrow}^\dagger a_{- k + \frac{q}{2} \downarrow}^\dagger a_{- k' + \frac{q}{2} \downarrow} a_{k' + \frac{q}{2} \uparrow} \notag \\
 & + \frac{1}{N} \sum_{k, k', q} V_b(k, k') b_{k + \frac{q}{2} \uparrow}^\dagger b_{- k + \frac{q}{2} \downarrow}^\dagger b_{- k' + \frac{q}{2} \downarrow} b_{k' + \frac{q}{2} \uparrow},
 \label{eq:model}
\end{align}
where $a_{k s}$ and $b_{k s}$ are the annihilation operators of electrons with spin $s = \uparrow, \downarrow$ on the sublattices $a$ and $b$, respectively.
The wave vector $k$ is directed to the crystallographic \textit{c} axis.

The first and second terms are the inter-sublattice and intra-sublattice hopping terms including the chemical potential $\mu$, respectively.
The kinetic energy $\varepsilon(k)$ and $\varepsilon'(k)$ are obtained by taking into account the nearest- and next-nearest-neighbor hoppings,
\begin{align}
 \varepsilon(k) &= - 2 t_1 \cos\frac{k}{2}, \\
 \varepsilon'(k) &= - 2 t_2 \cos k.
\end{align}
The crystal structure and hopping integrals, $t_1$ and $t_2$, are illustrated in Fig.~\ref{fig:1d-zigzag}.

The third term is a sublattice-dependent ASOC which originates from the violation of local inversion symmetry~\cite{Yanase2014}.
The $g$ vector is approximated as $\bm{g}(k) = \sin k \hat{z}$.
We choose the crystallographic \textit{a} axis as the quantization axis of the spin, namely, $\hat{z} = \hat{a}$. 

The fourth term expresses the molecular field of magnetic monopole, dipole, and quadrupole order.
This term causes various superconducting phenomena, which are demonstrated in this paper.
We assume that the Neel temperature $T_{\text{N}}$ is much larger than the superconducting transition temperature $T_{\text{C}}$.
In this situation, the fluctuation of multipole order is ignorable below $T_{\text{C}}$.
Effects of superconductivity on the magnetic order are also ignorable because the energy scale of superconductivity is much smaller than the magnetic interaction energy.
Therefore, our assumption for fixed  magnetic order is justified.

In order to study superconductivity in this system, we introduce an attractive interaction by the fifth and sixth terms in Eq.~\eqref{eq:model}, where $N$ is the number of sites in each sublattice.
For simplicity, we assume $s$-wave superconductivity by adopting the momentum-independent pairing interaction,
\begin{equation}
 V_a(k, k') = V_b(k, k') = - V.
\end{equation}
Although the spin-triplet $p$-wave order parameter is induced by the ASOC through either attractive or repulsive interaction in the $p$-wave channel, we neglect the $p$-wave order parameter.
It has been shown that the admixed $p$-wave component does not change the phase diagram unless the $p$-wave attractive interaction is comparable to or larger than the $s$-wave interaction~\cite{Yoshida2014}.

The purpose of this paper is to clarify exotic superconducting phases stabilized by the spin-orbit coupling and magnetic multipole order.
For this purpose, we treat a ``deep'' zigzag chain $t_1 / t_2 < 1$ and assume a moderate ASOC $\alpha / t_2 = 0.4$ so that the ASOC plays important roles.
The attractive interaction is chosen to be $V / t_2 = 1.5$ unless explicitly mentioned otherwise.

\subsection{Mean-field theory}
\label{sec:mf_theory}
Second, we investigate the superconducting state by means of mean-field (MF) theory. 
The interaction terms are approximated as follows: 
\begin{align}
 & - \frac{V}{N} \sum_{k, k', q} a_{k + \frac{q}{2} \uparrow}^\dagger a_{- k + \frac{q}{2} \downarrow}^\dagger a_{- k' + \frac{q}{2} \downarrow} a_{k' + \frac{q}{2} \uparrow} + (a \to b) \notag \\
 \simeq& \sum_{k} [ \Delta_a^* a_{- k + \frac{q}{2} \downarrow} a_{k + \frac{q}{2} \uparrow} + \text{h.c.} ] + \frac{N}{V} |\Delta_a|^2 + (a \to b),
\end{align}
by introducing the order parameter 
\begin{equation}
 \begin{split}
  \Delta_a &= - \frac{V}{N} \sum_{k'} \langle a_{- k' + \frac{q}{2} \downarrow} a_{k' + \frac{q}{2} \uparrow} \rangle, \\
  \Delta_b &= - \frac{V}{N} \sum_{k'} \langle b_{- k' + \frac{q}{2} \downarrow} b_{k' + \frac{q}{2} \uparrow} \rangle.
 \end{split}
 \label{eq:order-parameter}
\end{equation}
Thus, in this paper we assume a single-$q$ state.
The condensation energy is optimized with respect to the center-of-mass momentum $q$ of Cooper pairs.
In the BCS state and PDW state, $q = 0$ as we introduce in Sec.~\ref{sec:order_parameter}.
We also examine the $q \neq 0$ state corresponding to the FFLO state~\cite{FF, LO}.
The order parameters of the superconducting states are summarized in Sec.~\ref{sec:order_parameter}.

We here describe the MF Hamiltonian in a matrix form.
We define $k_+ \equiv k + \frac{q}{2}$, $k_- \equiv - k + \frac{q}{2}$, and the vector operator
\begin{equation}
 \hat{C}_{k}^\dagger \equiv (a_{k_+ \uparrow}^\dagger, a_{k_+ \downarrow}^\dagger, b_{k_+ \uparrow}^\dagger, b_{k_+ \downarrow}^\dagger, a_{k_- \uparrow}, a_{k_- \downarrow}, b_{k_- \uparrow}, b_{k_- \downarrow}).
\end{equation}
Then, we obtain
\begin{equation}
 {\cal H}_{\text{MF}} = \frac{1}{2} \sum_{k} \hat{C}_k^\dagger \hat{H}_8(k) \hat{C}_k + W_0,
\end{equation}
with
\begin{equation}
 W_0 = - \sum_{k} 2 [ \varepsilon'\left(k_- \right) - \mu ] + \frac{N}{V} |\Delta_a|^2 + \frac{N}{V} |\Delta_b|^2.
\end{equation}
The explicit form of the $8 \times 8$ matrix $\hat{H}_8(k)$ is given by 
\begin{equation}
 \hat{H}_8(k) = \begin{pmatrix}
                 \hat{H}_4(k_+) & \hat{\Delta}_4 \\
                 \hat{\Delta}_4^\dagger & - \hat{H}_4^{\text{T}}(k_-)
                \end{pmatrix},
                \label{eq:Hamiltonian}
\end{equation}
where
\begin{align}
 \hat{H}_4(k_\pm) &= \begin{pmatrix}
                      \hat{H}_2^{(a)}(k_\pm) - \mu \hat{\sigma}_0 & \varepsilon(k_\pm) \hat{\sigma}_0 \\
                      \varepsilon(k_\pm) \hat{\sigma}_0 & \hat{H}_2^{(b)}(k_\pm) - \mu \hat{\sigma}_0
                     \end{pmatrix}, \label{eq:Hamiltonian_4} \\
 \hat{\Delta}_4 &= \begin{pmatrix}
                    0 & \Delta_a & 0 & 0 \\
                    - \Delta_a & 0 & 0 & 0 \\
                    0 & 0 & 0 & \Delta_b \\
                    0 & 0 & - \Delta_b & 0
                   \end{pmatrix}, \\
 \hat{H}_2^{(l)}(k_\pm) &= \begin{cases}
                            \varepsilon'(k_\pm) \hat{\sigma}_0 + \alpha \sin k_\pm \hat{\sigma}_z - \bm{h}_a \cdot \bm{\hat{\sigma}} & (l = a) \\
                            \varepsilon'(k_\pm) \hat{\sigma}_0 - \alpha \sin k_\pm \hat{\sigma}_z - \bm{h}_b \cdot \bm{\hat{\sigma}} & (l = b).
                           \end{cases}
\end{align}

We carry out Bogoliubov transformation with using the unitary matrix $\hat{U}_8(k)$:
\begin{align}
 {\cal H}_{\text{MF}} &= \frac{1}{2} \sum_{k} \underbrace{\hat{C}_k^\dagger \hat{U}_8(k)}_{\hat{\Gamma}_k^\dagger} \underbrace{\hat{U}_8^\dagger(k) \hat{H}_8(k) \hat{U}_8(k)}_{\hat{E}_8(k)} \underbrace{\hat{U}_8^\dagger(k) \hat{C}_k}_{\hat{\Gamma}_k} + W_0 \notag \\
 &= \frac{1}{2} \sum_{k} \hat{\Gamma}_k^\dagger \hat{E}_8(k) \hat{\Gamma}_k + W_0,
\end{align}
where $\hat{E}_8(k)$ is a diagonal matrix,
\begin{equation}
 \hat{E}_8(k) = \begin{pmatrix}
                 \hat{E}_4(k) & \hat{0} \\
                 \hat{0} & - \hat{E}_4(k)
                \end{pmatrix}.
\end{equation}
From Eq.~\eqref{eq:order-parameter}, the order parameters are obtained by
\begin{align}
 \Delta_a &= - \frac{V_a}{N} \sum_{k'} \left\langle \left[ \hat{\Gamma}_{k'}^\dagger \hat{U}_8^\dagger(k') \right]_6 \left[ \hat{U}_8(k') \hat{\Gamma}_{k'} \right]_1 \right\rangle \notag \\
 &= - \frac{V_a}{N} \sum_{k'} \sum_{n = 1}^{8} \left[ \hat{U}_8^\dagger(k') \right]_{n 6} \left[ \hat{U}_8(k') \right]_{1 n} f\left( \left[ \hat{E}_8(k') \right]_{n n} \right), \label{eq:gap-equation_a} \\
 \Delta_b &= - \frac{V_a}{N} \sum_{k'} \sum_{n = 1}^{8} \left[ \hat{U}_8^\dagger(k') \right]_{n 8} \left[ \hat{U}_8(k') \right]_{3 n} f\left( \left[ \hat{E}_8(k') \right]_{n n} \right), \label{eq:gap-equation_b}
\end{align}
where $f(E)$ is the Fermi distribution function.
Equations \eqref{eq:gap-equation_a} and \eqref{eq:gap-equation_b} are MF gap equations to be solved numerically.

The Bogoliubov quasiparticle operator $\hat{\Gamma}_k^\dagger$ and energy $\hat{E}_4(k)$ are expressed with using the indices $(s, l)$, where $s$ represents the pseudospin $s = \uparrow, \downarrow$ and $l$ is the pseudo-sublattice index $l = a, b$:
\begin{gather}
 \hat{\Gamma}_k^\dagger = (\gamma_{k \uparrow a}^\dagger, \gamma_{k \downarrow a}^\dagger, \gamma_{k \uparrow b}^\dagger, \gamma_{k \downarrow b}^\dagger,
 \gamma_{- k \uparrow a}, \gamma_{- k \downarrow a}, \gamma_{- k \uparrow b}, \gamma_{- k \downarrow b}), \\
 \hat{E}_4(k) = \begin{pmatrix}
                 E_{k \uparrow a} & 0 & 0 & 0 \\
                 0 & E_{k \downarrow a} & 0 & 0 \\
                 0 & 0 & E_{k \uparrow b} & 0 \\
                 0 & 0 & 0 & E_{k \downarrow b}
                \end{pmatrix}.
\end{gather}
Then, the MF Hamiltonian ${\cal H}_{\text{MF}}$ and free energy $\Omega$ are obtained as
\begin{align}
 {\cal H}_{\text{MF}} &= \sum_{k, s, l} E_{k s l} \left( \gamma_{k s l}^\dagger \gamma_{k s l} - \frac{1}{2} \right) + W_0, \\
 \Omega &= - \frac{1}{\beta} \sum_{k, s, l} \left\{ \ln\left( 1 + e^{- \beta E_{k s l}} \right) + \frac{\beta E_{k s l}}{2} \right\} + W_0, \label{eq:free_energy}
\end{align}
where $\beta = 1 / T$ is the inverse temperature.

\section{Magnetic multipole order and exotic superconductivity}

\subsection{Magnetic and electronic structure in magnetic multipole states}
\label{sec:non-interacting}
We investigate the superconductivity in three magnetic multipole states: monopole, dipole, and quadrupole states.
Before going to the main issue, here we introduce the magnetic structure corresponding to the multipole order.
The symmetry protection on the single-particle band structure is also clarified.
Later we attribute the origin of exotic superconductivity to the unusual band structure.

First, we illustrate the magnetic structure in Fig.~\ref{fig:magnetic_structure}.
When the magnetic moment is ``antiferromagnetic'' in the unit cell and directed along the $x$ axis, two antiferromagnetic moments are regarded as a magnetic monopole [Fig.~\ref{fig:magnetic_structure}, panel (M)].
On the other hand, when the antiferromagnetic moment is parallel to the $z$ axis, a magnetic quadrupole moment is induced in the unit cell [Fig.~\ref{fig:magnetic_structure}, panel (Q)].
It has been shown that the magnetic quadrupole order is stabilized by the sublattice-dependent ASOC~\cite{Hayami2015}.
Indeed, the magnetic structure in 1-2-10 compounds resembles magnetic quadrupole order~\cite{Khalyavin2010, Tanida2011, Muro2011, Kato2011}.
This magnetic structure is also induced by the electric field applied along the \textit{c} axis as a result of the magnetoelectric effect~\cite{Yanase2014}.
The magnetic monopole and quadrupole are odd-parity multipoles leading to the spontaneous global inversion symmetry breaking.
Furthermore, we also examine the conventional ``ferromagnetic'' order which is called magnetic dipole order in this paper [Fig.~\ref{fig:magnetic_structure}, panel (D)].
The crystal structure of ferromagnetic SCs UGe$_2$~\cite{Saxena2000}, URhGe~\cite{Aoki2001}, and UCoGe~\cite{Huy2007} is composed of coupled zigzag chains~\cite{Aoki2012}.
Thus, our study may be relevant to these ferromagnetic SCs.

\begin{figure}[htbp]
 \centering
 \includegraphics[width=80mm, clip]{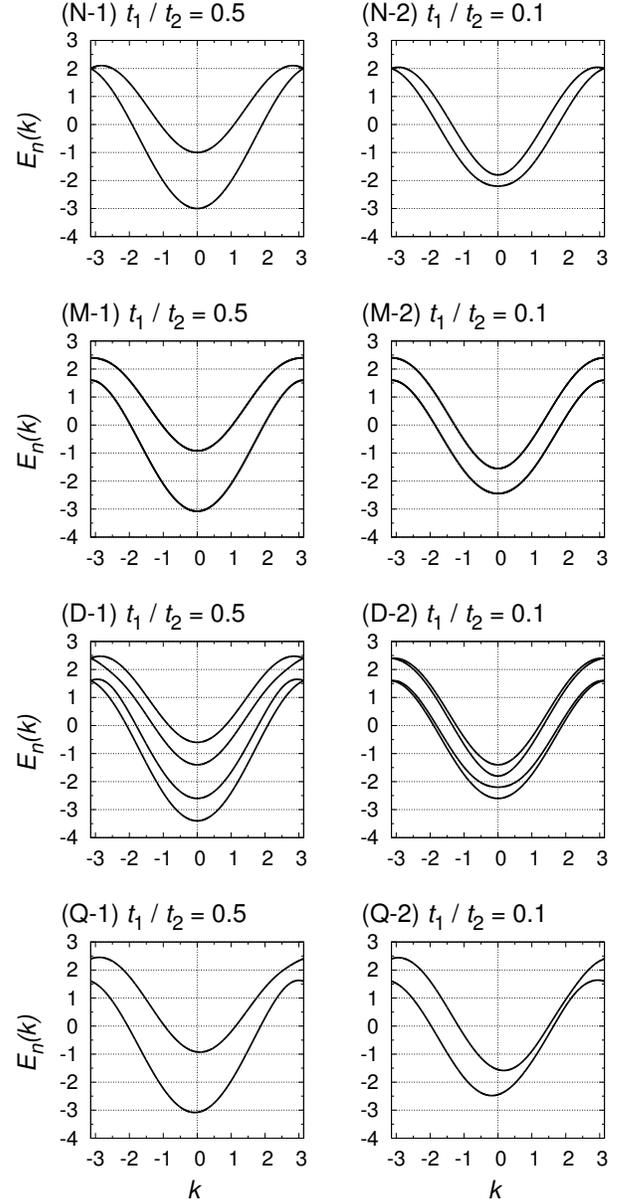}
 \caption{Band structure of 1D zigzag chain in (N) normal state, (M) magnetic monopole state, (D) magnetic dipole state, and (Q) magnetic quadrupole state. The left panels show the results for $t_1 / t_2 = 0.5$, while $t_1 / t_2 = 0.1$ in the right panels. In (M), (D), and (Q), we choose the molecular field $h = 0.4$ and $h^{\text{AF}} = 0.4$.}
 \label{fig:energy_band(N)}
\end{figure}

Next, we clarify the single-particle energy spectrum.
The band structure is obtained by the normal part Hamiltonian, which is expressed by using the vector operator $\hat{D}_k^\dagger = (a_{k \uparrow}^\dagger, a_{k \downarrow}^\dagger, b_{k \uparrow}^\dagger, b_{k \downarrow}^\dagger)$,
\begin{equation}
 H^{(0)} = \sum_{k} \hat{D}_k^\dagger \hat{H}_4(k) \hat{D}_k.
\end{equation}
Without any loss of generality, we choose the chemical potential $\mu$ to be zero in $\hat{H}_4(k)$ [Eq.~\eqref{eq:Hamiltonian_4}].
The itinerant magnetic multipole states are studied by taking into account the molecular field $\bm{h}_a$ and $\bm{h}_b$ as follows:
\begin{equation}
 (\bm{h}_a, \bm{h}_b) = \begin{cases}
                         (h^{\text{AF}} \hat{x}, - h^{\text{AF}} \hat{x}) & \text{in (M)onopole order} \\
                         (h \hat{y}, h \hat{y}) & \text{in (D)ipole order} \\
                         (h^{\text{AF}} \hat{z}, - h^{\text{AF}} \hat{z}) & \text{in (Q)uadrupole order.}
                        \end{cases}
\end{equation}
Then we show the energy band in Fig.~\ref{fig:energy_band(N)}.
In the absence of the magnetic multipole order, namely, $(\bm{h}_a, \bm{h}_b) = (0, 0)$, two bands are expressed by the following dispersion relation [Fig.~\ref{fig:energy_band(N)}, panels (N-1) and (N-2)]:
\begin{equation}
 E_n(k) = \varepsilon'(k) \pm \sqrt{\varepsilon(k)^2 + \alpha^2 \sin^2 k}.
\end{equation}
Each band has a twofold degeneracy which arises from the spin and sublattice degrees of freedom entangled by the sublattice-dependent ASOC.
This electronic structure is similar to the bilayer Rashba system studied in the previous study~\cite{Maruyama2012}.
On the other hand, we obtain the dispersion relation in the magnetic multipole states,
\begin{equation}
 E_n(k) = \begin{cases}
         \varepsilon'(k) \pm \sqrt{\varepsilon(k)^2 + \alpha^2 \sin^2 k + \left( h^{\text{AF}} \right)^2} & \text{in (M)} \\
         \varepsilon'(k) \pm \sqrt{\left[ \varepsilon(k) \pm h \right]^2 + \alpha^2 \sin^2 k} & \text{in (D)} \\
         \varepsilon'(k) \pm \sqrt{\varepsilon(k)^2 + \left( \alpha \sin k - h^{\text{AF}} \right)^2} & \text{in (Q).}
        \end{cases}
\end{equation}

Table~\ref{tab:band_structure} shows two main features of the band structure: (i) symmetry with respect to the inversion of momentum, $k \to - k$, and (ii) twofold degeneracy.
Below, we explain these features in terms of symmetry in multipole states.

\begin{table}[htbp]
 \centering
 \caption{Band structure in the magnetic multipole states.}
 \label{tab:band_structure}
 \setlength{\tabcolsep}{8pt}
 \begin{tabular}{r@{\hspace{20pt}}cc} \hline\hline
             & (i) Symmetry & (ii) Twofold degeneracy \\ \hline
  Monopole   & yes          & yes                     \\
  Dipole     & yes          & no                      \\
  Quadrupole & no           & yes                     \\ \hline\hline
 \end{tabular}
\end{table}

First, in the magnetic monopole state, the collinear antiferromagnetic order spontaneously breaks the inversion symmetry (${\cal P}$ symmetry) as well as the time-reversal symmetry (${\cal T}$ symmetry) in spite of the globally centrosymmetric crystal structure.
However the combined ${\cal PT}$ symmetry is preserved since the normal part Hamiltonian $H^{(0)}$ is invariant under the successive operations of time-reversal and spatial inversion.
This combined operation satisfies $({\cal PT})^2 = - 1$ which ensures a twofold degeneracy in the band structure as proved by the Kramers theorem.
Furthermore, the system transforms under the twofold rotation as follows:
\begin{align}
 {\cal R}_x^\pi \hat{H}_4(k) ({\cal R}_x^\pi)^{- 1} &= \hat{H}_4(- k), \label{eq:twofold_rotation_x} \\
 {\cal R}_z^\pi \hat{H}_4(k) ({\cal R}_z^\pi)^{- 1} &= \hat{H}_4(- k) \quad \text{in (M)}. \label{eq:twofold_rotation_z}
\end{align}
From Eq.~\eqref{eq:twofold_rotation_x} or \eqref{eq:twofold_rotation_z}, we understand the symmetric energy dispersion, $E_n(k) = E_n(- k)$.
Second, in the magnetic quadrupole state, the band structure preserves a twofold degeneracy owing to the very same reason as the monopole state.
However, the quadrupole state is neither invariant under the twofold rotation nor the mirror reflection with respect to the $zx$ plane which transforms the wave number $k$ to $- k$:
\begin{align}
 {\cal R}_x^\pi \hat{H}_4(k) ({\cal R}_x^\pi)^{- 1} &\neq \hat{H}_4(- k),\\
 {\cal R}_z^\pi \hat{H}_4(k) ({\cal R}_z^\pi)^{- 1} &\neq \hat{H}_4(- k),\\
 {\cal M}_{zx} \hat{H}_4(k) {\cal M}_{zx}^{- 1} &\neq \hat{H}_4(- k) \quad \text{in (Q)}.
\end{align}
Thus, all the symmetries protecting the symmetric band structure are broken, and indeed, the band structure is asymmetric as shown in Fig.~\ref{fig:energy_band(N)}, panels (Q-1) and (Q-2).
Finally, the band structure is symmetric in the magnetic dipole state since the ferromagnetic order preserves the ${\cal P}$ symmetry.
Because of the violation of the ${\cal T}$ symmetry the combined ${\cal PT}$ symmetry is broken, and therefore, the twofold degeneracy is lifted. 

We furthermore show the symmetry protection on the additional degeneracy at $k = \pm \pi$.
For example, two spinful bands are degenerate at $k = \pm \pi$ in the normal state.
This fourfold degeneracy is protected by the ${\cal PT}$ symmetry, inversion-glide symmetry ${\cal PG}_{yz}$, and mirror symmetry ${\cal M}_{xz}$.
We here prove the fourfold degeneracy at the inversion-glide-invariant momentum $k = \pm \pi$ from relations, $({\cal PG}_{yz})^2 = - 1$, $\{ {\cal PG}_{yz}, {\cal PT} \} = 0$, and $\{ {\cal PG}_{yz}, {\cal M}_{xz} \} = 0$~\cite{QiFeng2016}.
Because of the inversion-glide symmetry, the normal part Hamiltonian at $k = \pm \pi$ is block diagonalized and decomposed into the $\pm i$ subsectors.
The ${\cal PT}$ symmetry is preserved in each subsector as ensured by the anticommutation relation between ${\cal PG}_{yz}$ and ${\cal PT}$.
Thus, Kramers pairs are formed in each subsector.
The anticommutation relation between ${\cal PG}_{yz}$ and ${\cal M}_{xz}$ ensures that a Kramers pair in the $i$ subsector is degenerate with another Kramers pair in the $- i$ subsector.
Thus, the fourfold degeneracy is protected by symmetry.
The mirror symmetry is broken in the monopole and quadrupole states, while the ${\cal PT}$ symmetry is broken in the dipole state.
Therefore, the fourfold degeneracy is lifted in the multipole states. 

Additional degeneracy is also seen at $k = \pm \pi$ in the dipole state because the normal part Hamiltonian preserves the magnetic-glide symmetry ${\cal TG}_{yz}$ which is a combined symmetry of the glide symmetry and the time-reversal symmetry.
This antiunitary symmetry ensures the extended Kramers theorem proving the degenerate single-particle states at $k = \pm \pi$.
The twofold degeneracy is also protected in the magnetic dipole state with $\bm{h}$ along the $x$ axis.
Then, the magnetic-screw symmetry ${\cal TS}_y^\pi$ protects the degeneracy at $k = \pm \pi$.

\subsection{Superconductivity}
\label{sec:order_parameter}
We here summarize the order parameter of three superconducting states which may be stabilized in our model: the BCS state, the PDW state, and the FFLO state.
In the conventional BCS state, Cooper pairs have the zero in-plane center-of-mass momentum, that is, $q = 0$.
The order parameter is uniform between sublattices, $(\Delta_a, \Delta_b) = (\Delta, \Delta)$.
The center-of-mass momentum is also zero in the PDW state.
The sign of the order parameter, however, changes between sublattices, $(\Delta_a, \Delta_b) = (\Delta, - \Delta)$.
In the FFLO state, the center-of-mass momentum of Cooper pairs is finite (i.e., $q \neq 0$).
In the symmetry considered in this paper, the Cooper pair condensation occurs at a single $q$, although the double-$q$ state is stable in the conventional FFLO state~\cite{Matsuda_FFLO}.
Therefore, in real space the order parameter is expressed as $\Delta(y) = \Delta e^{i q y}$, which is usually called ``Fulde-Ferrell state''~\cite{FF} or ``helical state''~\cite{Edelstein, Dimitrova, Samokhin,Agterberg}.
As is the case in the BCS state, the order parameter is uniform in sublattices, $(\Delta_a, \Delta_b) = (\Delta, \Delta)$.

\section{BCS state robust against magnetic monopole order}
\label{sec:monopole}
\begin{figure}[htbp]
 \centering
 \includegraphics[width=75mm, clip]{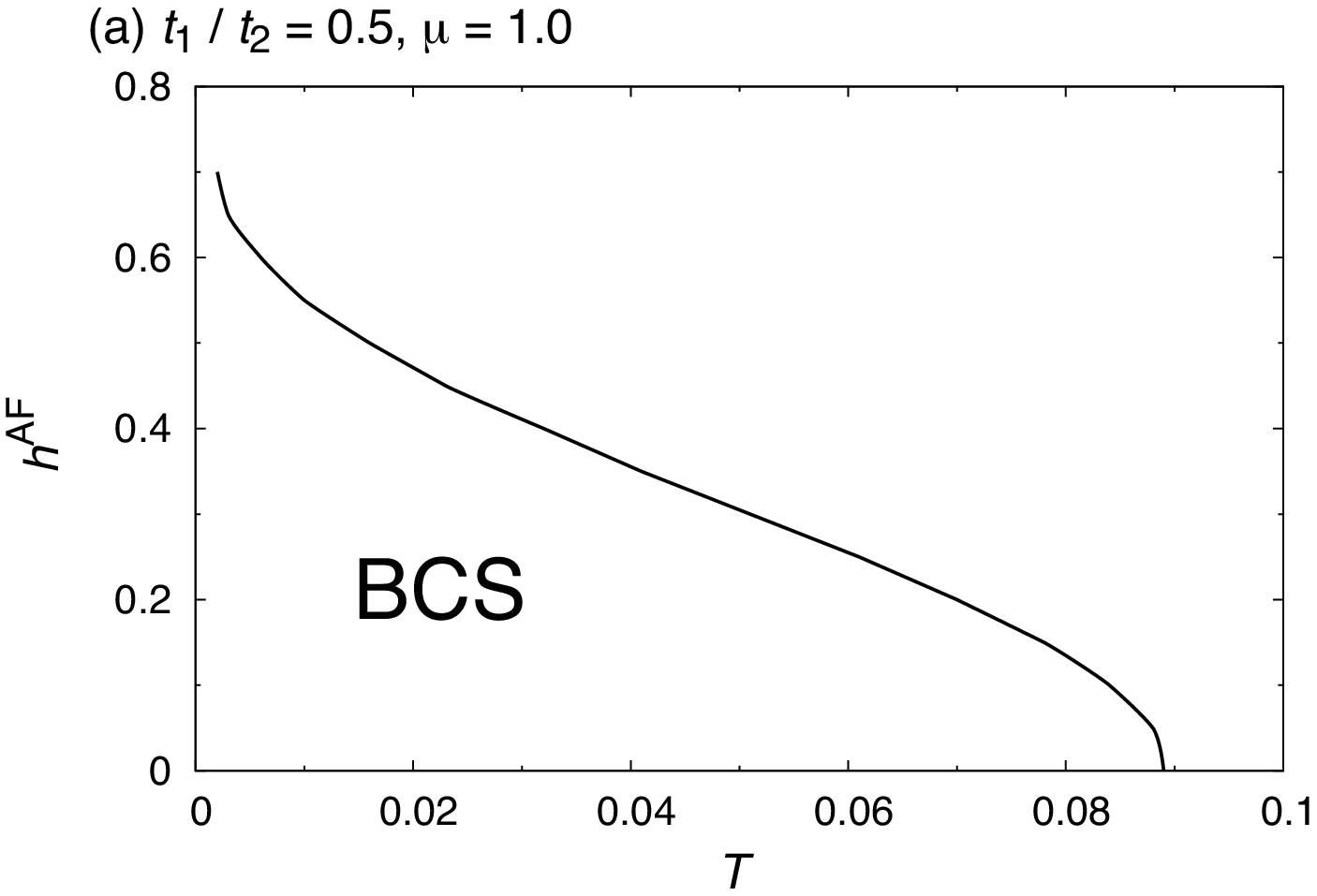}
 \includegraphics[width=75mm, clip]{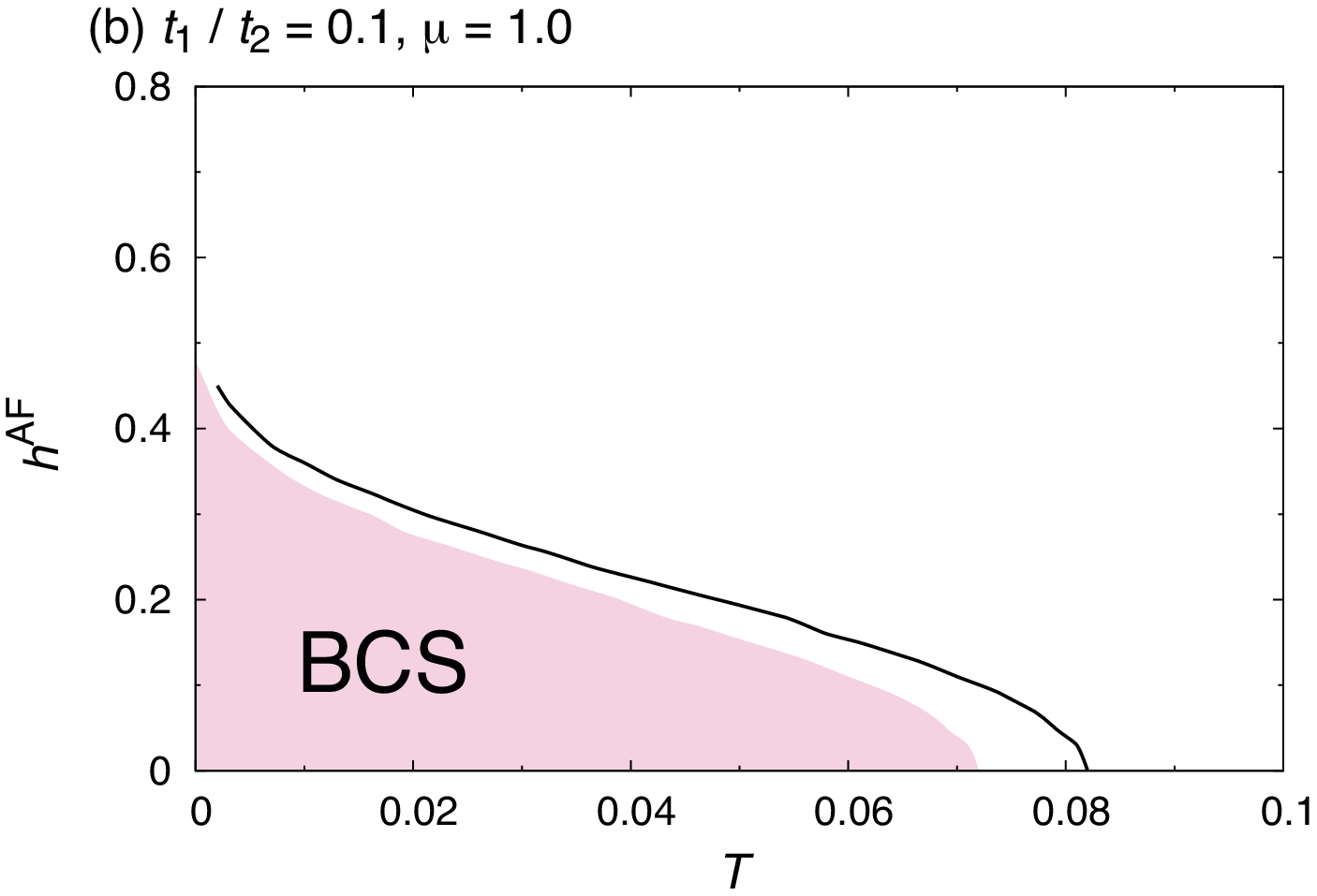}
 \includegraphics[width=75mm, clip]{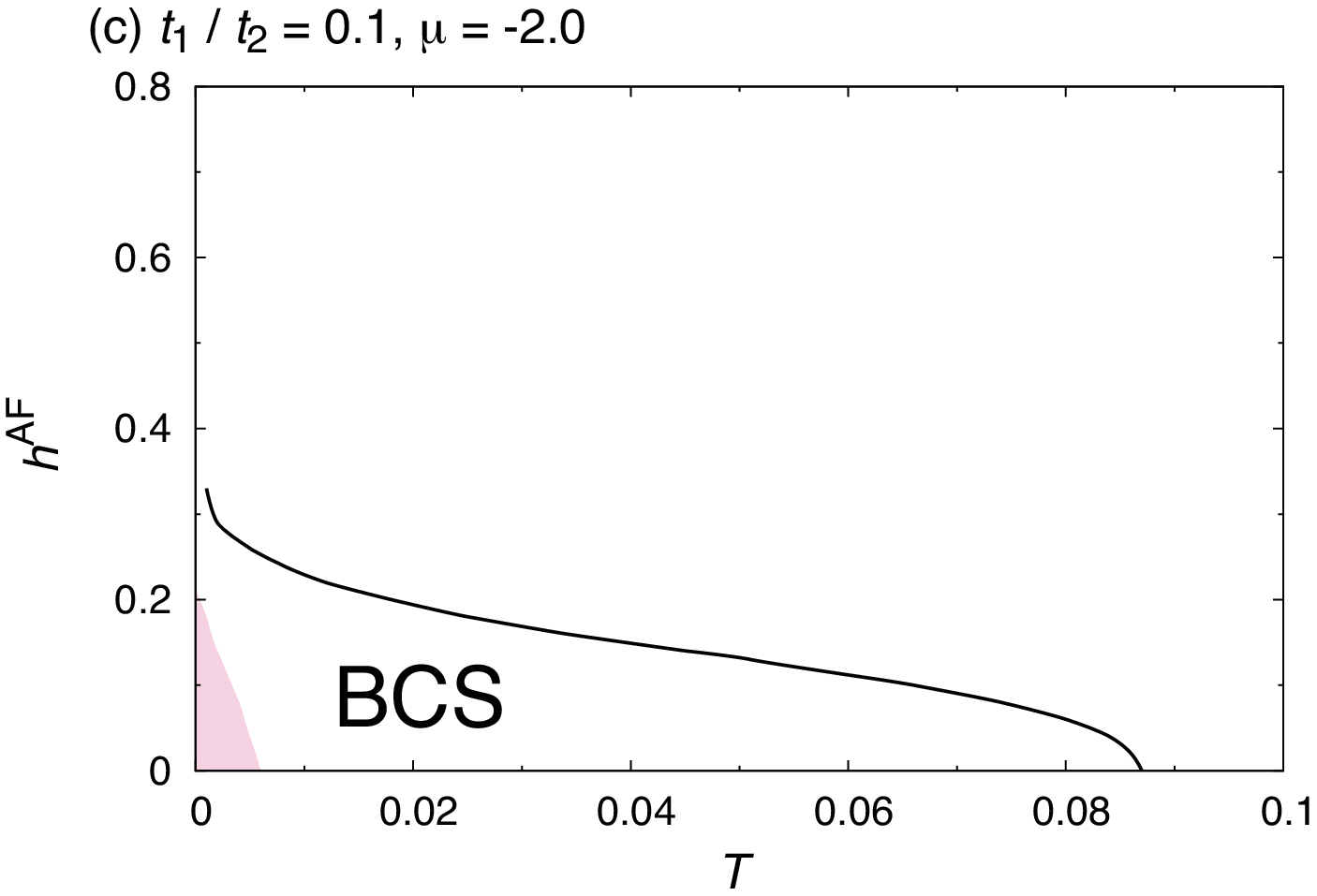}
 \caption{$T$-$h^{\text{AF}}$ phase diagram in the magnetic monopole state for (a) $t_1 / t_2 = 0.5, \mu = 1$, (b) $t_1 / t_2 = 0.1, \mu = 1$, and (c) $t_1 / t_2 = 0.1, \mu = - 2$. The BCS state is stable in the whole superconducting phase. In the pink shaded area the PDW state is metastable.}
 \label{fig:T-h_phase_MP}
\end{figure}

First we discuss the superconductivity coexisting with magnetic monopole order [see Fig.~\ref{fig:magnetic_structure}, panel (M)].
Figure~\ref{fig:T-h_phase_MP} shows $T$-$h^{\text{AF}}$ phase diagrams for several sets of parameters.
It is shown that the conventional BCS state is stable in the whole phase diagram independently of the parameters $\mu$ and $t_1 / t_2$.
For $t_1 / t_2 = 0.5$ [Fig.~\ref{fig:T-h_phase_MP}(a)], the PDW state is not even a metastable state.
On the other hand, for $t_1 / t_2 = 0.1$ [Figs.~\ref{fig:T-h_phase_MP}(b) and \ref{fig:T-h_phase_MP}(c)] the PDW state is metastable in a certain parameter regime, indicated by a negative condensation energy whose absolute value is smaller than that of the BCS state.
The phase diagram for $\mu = 2$ is similar to that for $\mu = 1$, although we do not show in Fig.~\ref{fig:T-h_phase_MP}.
Thus, the BCS state is stable in the monopole state, irrespective of the number of Fermi surfaces.
We confirmed that the in-plane center-of-mass momentum $q$ of the Cooper pair is zero in the whole parameter region.

\section{PDW state by magnetic dipole order}
\label{sec:dipole}
Second, we study the superconductivity in the magnetic dipole state [see Fig.~\ref{fig:magnetic_structure}, panel (D)]. This situation is realized when the superconductivity occurs in the ferromagnetic metal. Indeed, such ferromagnetic superconductivity occurs in uranium-based heavy-fermion SCs UGe$_2$~\cite{Saxena2000}, URhGe~\cite{Aoki2001}, and UCoGe~\cite{Huy2007}, which have a zigzag crystal structure~\cite{Aoki2012}. Although the magnetic dipole moment along the $y$ axis is assumed, the $x$ axis is equivalent to the $y$ axis in the spin space since we consider a purely 1D model. Note that both the $x$ and $y$ axes are perpendicular to the $g$ vector.

\subsection{Phase diagram}
\label{sec:dipoleA}
\begin{figure}[htbp]
 \centering
 \includegraphics[width=75mm, clip]{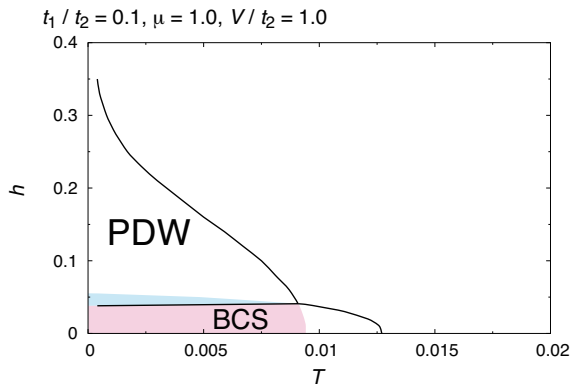}
 \caption{$T$-$h$ phase diagram in the magnetic dipole state for $t_1 / t_2 = 0.1$ and $\mu = 1$. We assume the attractive interaction $V / t_2 = 1.0$. In the pink (cyan) shaded area the PDW (BCS) state is a metastable state.}
 \label{fig:T-h_phase_DP}
\end{figure}

Figure~\ref{fig:T-h_phase_DP} shows the $T$-$h$ phase diagram for $t_1 / t_2 = 0.1$.
The BCS state is stable in a weakly polarized region (small $h$ region), while the PDW state is stable in a large parameter range with high spin polarization (large $h$ region).
The phase boundary of the BCS and PDW states is the first-order phase transition line.
This phase diagram is similar to that obtained in the two-dimensional bilayer Rashba SCs~\cite{Yoshida2012}.

The mechanism of the PDW state in a spin polarized state has been discussed in Ref.~\onlinecite{Yoshida2012}.
When the inter-sublattice hopping is smaller than the spin-orbit coupling, a substantial condensation energy is gained in the PDW state although at zero effective magnetic field ($h = 0$) it is smaller than the condensation energy of the BCS state which gains the inter-sublattice Josephson coupling energy.
Because the paramagnetic depairing effect is suppressed in the PDW state by the spin-orbit coupling~\cite{Maruyama2012}, at large $h$ the PDW state may be more stable than the BCS state which is fragile against the paramagnetic effect. 

The zigzag chain is composed of the ``$a$ sublattice'' and the ``$b$ sublattice,'' and thus $t_1$ is the inter-sublattice hopping.
When the system has a small $t_1 / t_2$ and a moderate ASOC, the PDW state is stabilized in a large parameter regime as shown in Fig.~\ref{fig:T-h_phase_DP}.
As the inter-sublattice hopping $t_1 / t_2$ is increased, the PDW state is suppressed.
For our choice of parameters, the PDW state is not stable for $t_1 / t_2 > 0.7$.
Thus, in the zigzag chain the PDW state may be stable even at a moderate $t_1 / t_2$.
This is partly because the inter-sublattice coupling is represented by $\varepsilon(k_\text{F})$ rather than $t_1$, and $\varepsilon(k)$ disappears at $k = \pm \pi$.
As we mentioned in Sec.~\ref{sec:non-interacting}, the fourfold degeneracy at $k = \pm \pi$ is protected by the inversion-glide symmetry ${\cal PG}_{yz}$ and mirror symmetry.
This additional degeneracy comes from the sublattice degree of freedom.
Thus, the disappearance of $\varepsilon(\pm\pi)$ is ensured by the nonsymmorphic crystal symmetry.
When the Fermi momentum is close to $k = \pm \pi$, the PDW state is favored owing to a small $\varepsilon(k_\text{F})$.

As shown in Fig.~\ref{fig:T-h_phase_DP}, the PDW state may be stable at $\mu = 1$, where the four energy bands cross the Fermi level [see Fig.~\ref{fig:energy_band(N)}, panel (D-2)].
Similarly, the PDW state is stable when two or three energy bands have the Fermi surface.
We confirmed that the in-plane center-of-mass momentum $q$ of the Cooper pair is zero in any case.
When the chemical potential is in the vicinity of the band edge and only one band crosses the Fermi level, however, the PDW state is not stable.

\subsection{Topological superconductivity}
\label{sec:topological}
In this subsection, we show that the PDW state may be a 1D topological superconducting state specified by the winding number and the $\mathbb{Z}_2$ invariant.
A gauge transformation, $a_k^\dagger \to a_k^\dagger e^{i k / 2}$, is carried out so that the Bogoliubov-de Gennes (BdG) Hamiltonian is periodic in the Brillouin zone.
This unitary transformation is useful for the discussion of topological properties in nonsymmorphic systems~\cite{Shiozaki2015}.

First, we elucidate the winding number.
In a ferromagnetic state with magnetic moment along the $x$ or $y$ axis, the system is invariant under the magnetic mirror reflection which is a successive operation of time reversal ${\cal T} = i \hat{\sigma}_y {\cal K}$ and mirror reflection with respect to the $xy$ plane ${\cal M}_{xy} = i \hat{\sigma}_z$.
${\cal K}$ is the complex-conjugate operator.
Thus, the BdG Hamiltonian derived from Eq.~\eqref{eq:Hamiltonian} preserves the pseudo-time-reversal symmetry:
\begin{equation}
 {\cal T}'_8 \hat{H}_8(- k) {{\cal T}'_8}^\dagger = \hat{H}_8(k),
\end{equation}
where
\begin{equation}
 {\cal T}'_8 = \begin{pmatrix}
                {\cal T}' & \hat{0} \\
                \hat{0} & {\cal T}'^*
               \end{pmatrix},
\end{equation}
with ${\cal T}' = {\cal M}_{xy} {\cal T}$.
Furthermore, the particle-hole symmetry is implemented in the BdG Hamiltonian:
\begin{equation}
 {\cal C} \hat{H}_8(- k) {\cal C}^\dagger = - \hat{H}_8(k),
\end{equation}
where ${\cal C} = \tau_x {\cal K}$ and $\tau_x$ is the Pauli matrix in the particle-hole space.
Combining the pseudo-time-reversal symmetry with the particle-hole symmetry, we can define the chiral symmetry,
\begin{equation}
 \{ \Gamma, \hat{H}_8(k) \} = 0,
\end{equation}
with $\Gamma = - {\cal C} {\cal T}'_8$.
The chiral symmetry ensures that the 1D winding number
\begin{equation}
 \omega = \frac{1}{4 \pi i} \int_{- \pi}^{\pi} dk \Tr\left[ \hat{q}(k)^{- 1} \partial_k \hat{q}(k) - \hat{q}^\dagger(k)^{- 1} \partial_k \hat{q}^\dagger(k) \right]
\end{equation}
is a $\mathbb{Z}$ topological invariant~\cite{Sato2009, Sato2011, Yada2011, Schnyder2011, Tewari2012, Wong2012, Watanabe2015} when a finite gap is open.
The $4 \times 4$ matrix $\hat{q}(k)$ is obtained by carrying out a unitary transformation
\begin{equation}
 \hat{V} \hat{H}_8(k) \hat{V}^\dagger = \begin{pmatrix}
                                         \hat{0} & \hat{q}(k) \\
                                         \hat{q}^\dagger(k) & \hat{0}
                                        \end{pmatrix},
\end{equation}
where $\hat{V}$ is a unitary matrix which diagonalizes $\Gamma$~\cite{Wen}.

\begin{figure}[htbp]
 \centering
 \includegraphics[width=70mm, clip]{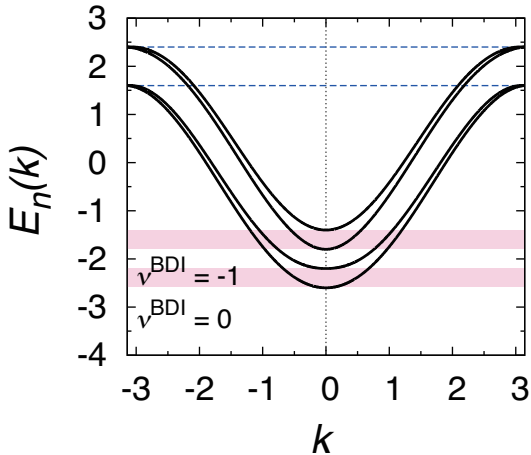}
 \caption{Chemical potential dependence of the winding number $\nu^{\text{BDI}}$ for $t_1 / t_2 = 0.1$, $h = 0.40$, and $\Delta = 0.01$. The winding number is nontrivial, $\nu^{\text{BDI}} = - 1$, when the chemical potential lies in the pink shaded region. The blue dashed lines represent the chemical potential at which the winding number is ill defined owing to the gap closing.}
 \label{fig:winding_num}
\end{figure}

The BdG Hamiltonian $\hat{H}_8(k)$ belongs to the symmetry class BDI because $({\cal T}'_8)^2 = + 1$ and ${\cal C}^2 = + 1$.
Therefore, the winding number $\omega$ is identified to be an integer topological invariant of the BDI class~\cite{Schnyder2008, Kitaev2009, Ryu2010}, $\nu^{\text{BDI}}$.
Figure~\ref{fig:winding_num} shows the chemical potential dependence of the winding number together with the energy bands shown in Fig.~\ref{fig:energy_band(N)}, panel (D-2).
We obtain a finite winding number, $\nu^{\text{BDI}} = - 1$, indicating topologically nontrivial properties when one or three bands cross the Fermi level.
Otherwise, the winding number is trivial, $\nu^{\text{BDI}} = 0$.

\begin{figure}[htbp]
 \centering
 \includegraphics[width=85mm, clip]{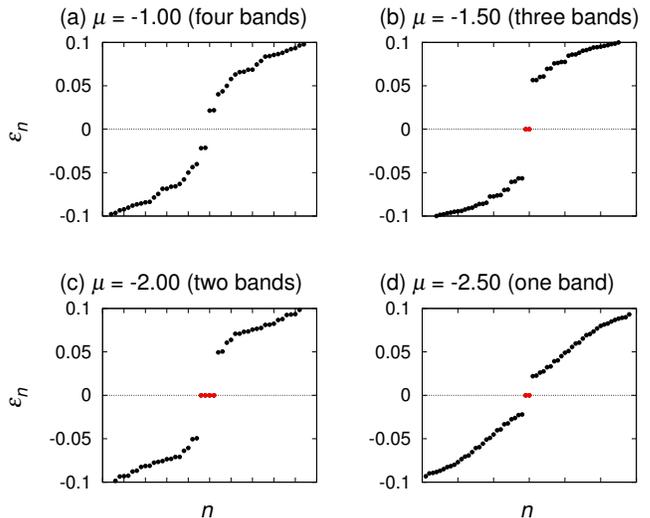}
 \caption{Energy spectra in the PDW state with open boundaries. We assume the ferromagnetic molecular field $h = 0.4$ and (a) $\mu = - 1.00$, (b) $\mu = - 1.50$, (c) $\mu = - 2.00$, and (d) $\mu = - 2.50$. 
The number of Fermi surface is 4, 3, 2, and 1. 
The other parameters are $t_1 / t_2 = 0.1$ and $(\Delta_a, \Delta_b) = (0.1, - 0.1)$. The red points indicate the Majorana end states. The number of Majorana states is doubled by the two boundaries.}
 \label{fig:edge_state}
\end{figure}

A nontrivial winding number may ensure the Majorana end state according to the index theorem~\cite{Sato2011}.
Indeed, Fig.~\ref{fig:edge_state} shows the Majorana end states.
The energy spectrum $\varepsilon_n$ is obtained in the open boundary condition, and the $n$-th energy eigenvalue is arranged in ascending order $\varepsilon_0 < \varepsilon_1 < \dotsb$.
We see the single Majorana end state protected by the nontrivial winding number $\nu^{\text{BDI}} = - 1$ in Figs.~\ref{fig:edge_state}(b) and \ref{fig:edge_state}(d).

This single Majorana end state is robust against perturbations, even when the magnetic mirror symmetry is broken.
Indeed, the PDW state with $\nu^{\text{BDI}} = - 1$ is a strong topological SC specified by the $\mathbb{Z}_2$ invariant in the D class~\cite{Schnyder2008, Kitaev2009, Ryu2010}.
The parity of the winding number is equivalent to the $\mathbb{Z}_2$ invariant, $\nu$, which is explicitly expressed by the Berry phase
\begin{equation}
 W[C] = \frac{1}{2 \pi} \sum_{n \in \text{occupied}} \oint_C dk \, i \langle u_n(k) | \partial_k | u_n(k) \rangle. 
\end{equation}
$C$ represents a time-reversal-invariant (TRI) closed path in the Brillouin zone, and $| u_n(k) \rangle$ is an eigenstate of $\hat{H}_8(k)$.
Since the BdG Hamiltonian preserves the particle-hole symmetry, the Berry phase is quantized as $e^{2 \pi i W[C]} = \pm 1$~\cite{LiangQi2008}.
Since the TRI closed path $C = \{ k \in [-\pi : \pi) \}$ is unique in the 1D system, we have a single $\mathbb{Z}_2$ invariant $e^{2 \pi i W[C]} = (- 1)^{\nu}$.
In particular, the normal part Hamiltonian preserves the spatial inversion symmetry, ${\cal P} \hat{H}_4(k) {\cal P}^\dagger = \hat{H}_4(- k)$, and the parity of the gap function is odd, ${\cal P} \hat{\Delta}_4 {\cal P}^{\text{T}} = - \hat{\Delta}_4$, in the PDW state.
Then, the $\mathbb{Z}_2$ invariant has been evaluated as
\begin{equation}
 (- 1)^\nu = \prod_{n} \sgn E_n(\Gamma_1) \sgn E_n(\Gamma_2),
\end{equation}
where $\Gamma_1$ and $\Gamma_2$ are the TRI momenta, $\Gamma_1 = 0$ and $\Gamma_2 = \pi$~\cite{Sato2010}.
From this representation, the $\mathbb{Z}_2$ invariant is nontrivial when the odd number of bands cross the Fermi level.
This condition coincides with the situation with $\nu^{\text{BDI}} = - 1$.
Thus, the PDW state is identified to be a 1D $\mathbb{Z}_2$ topological SC in the D class.

An intuitive explanation for the topological superconductivity is obtained by looking at the band representation of the BdG Hamiltonian,
\begin{equation}
 \hat{U}(k)^\dagger \hat{H}_8(k) \hat{U}(k) \simeq \bigoplus_{n = 1}^{4} \begin{pmatrix}
                                                                          E_n(k) & \Delta_n(k) \\
                                                                          \Delta_n^*(k) & - E_n(- k)
                                                                         \end{pmatrix},
\end{equation}
where $\hat{U}(k) = \left( \begin{smallmatrix} \hat{U}_4(k) & \hat{0} \\ \hat{0} & \hat{U}^*_4(- k) \end{smallmatrix} \right)$, and $\hat{U}_4(k)$ is a unitary matrix which diagonalizes $\hat{H}_4(k)$.
The order parameter in the band basis approximately has the $p$-wave form, $\Delta_n(k) \sim \sin k$.
In this sense, the situation is similar to the Kitaev chain~\cite{Kitaev2001} for the spinless $p$-wave SC.
Although the superconductivity is induced by the conventional pairing interaction in the $s$-wave spin-singlet channel, the effective $p$-wave superconducting state similar to the Kitaev chain is realized by the inter-sublattice phase modulation in the order parameter. 

Topological superconducting phases in 1D noncentrosymmetric systems have been clarified theoretically~\cite{PhysRevLett.105.077001, Hui2015}, and recently experimental indications for the Majorana state have been obtained in semiconductors~\cite{Mourik2012, Rokhinson2012} and ferromagnetic atomic chains~\cite{Nadj-Perge602}.
In contrast to these systems requiring the inversion-symmetry breaking, our research proposes the centrosymmetric topological superconductivity caused by the spontaneously formed odd-parity PDW order parameter.

Now we briefly comment on the zero energy end states in Fig.~\ref{fig:edge_state}(c).
When the two bands cross the Fermi level, we see the two Majorana end states in spite of the trivial winding number and $\mathbb{Z}_2$ number, $\nu^{\text{BDI}} = \nu = 0$.
These end states may be protected by another symmetry.
However, the crystal symmetry other than the mirror symmetry is broken at the boundary.
Thus, we leave the topological protection of these end states for a future study. 

Finally, we propose two experimental tests to identify the PDW state.
(i) As shown above, the Majorana end state is generated at the end of the chain in the PDW state.
The Majorana end state may be recognized as a zero bias conductance peak of quasiparticle tunneling spectroscopy in a normal metal/SC junction~\cite{Tanaka_special}.
(ii) In the external magnetic field, vortices appear in the real 3D materials.
Then, the local quasiparticle density of states in the PDW state is quite different from that in the BCS state.
The zero-energy vortex bound state exists in the PDW state, although it is absent in the BCS state due to the Zeeman effect~\cite{Higashi2016}.
Therefore, the scanning tunneling microscopy/spectroscopy experiments may identify the PDW state by measuring the local density of states.

\section{FFLO state by magnetic quadrupole order}
\label{sec:quadrupole}
Finally we clarify the superconductivity in the magnetic quadrupole state [see Fig.~\ref{fig:magnetic_structure}, panel (Q)].
We assume $t_1 / t_2 = 0.5$ in what follows.

As we showed in Sec.~\ref{sec:non-interacting}, energy bands are asymmetric in the magnetic quadrupole state [see Fig.~\ref{fig:energy_band(N)}, panel (Q-1)] in sharp contrast to the normal and other multipole states.
Roughly speaking, the upper (lower) band is distorted into the positive (negative) momentum direction for parameters in Fig.~\ref{fig:energy_band(N)}.
This unusual band structure may stabilize an exotic superconducting state.
Indeed, we show that the FFLO state is stabilized even at zero magnetic field.

\subsection{$T$-$\mu$ phase diagram}
\begin{figure}[htbp]
 \centering
 \includegraphics[width=75mm, clip]{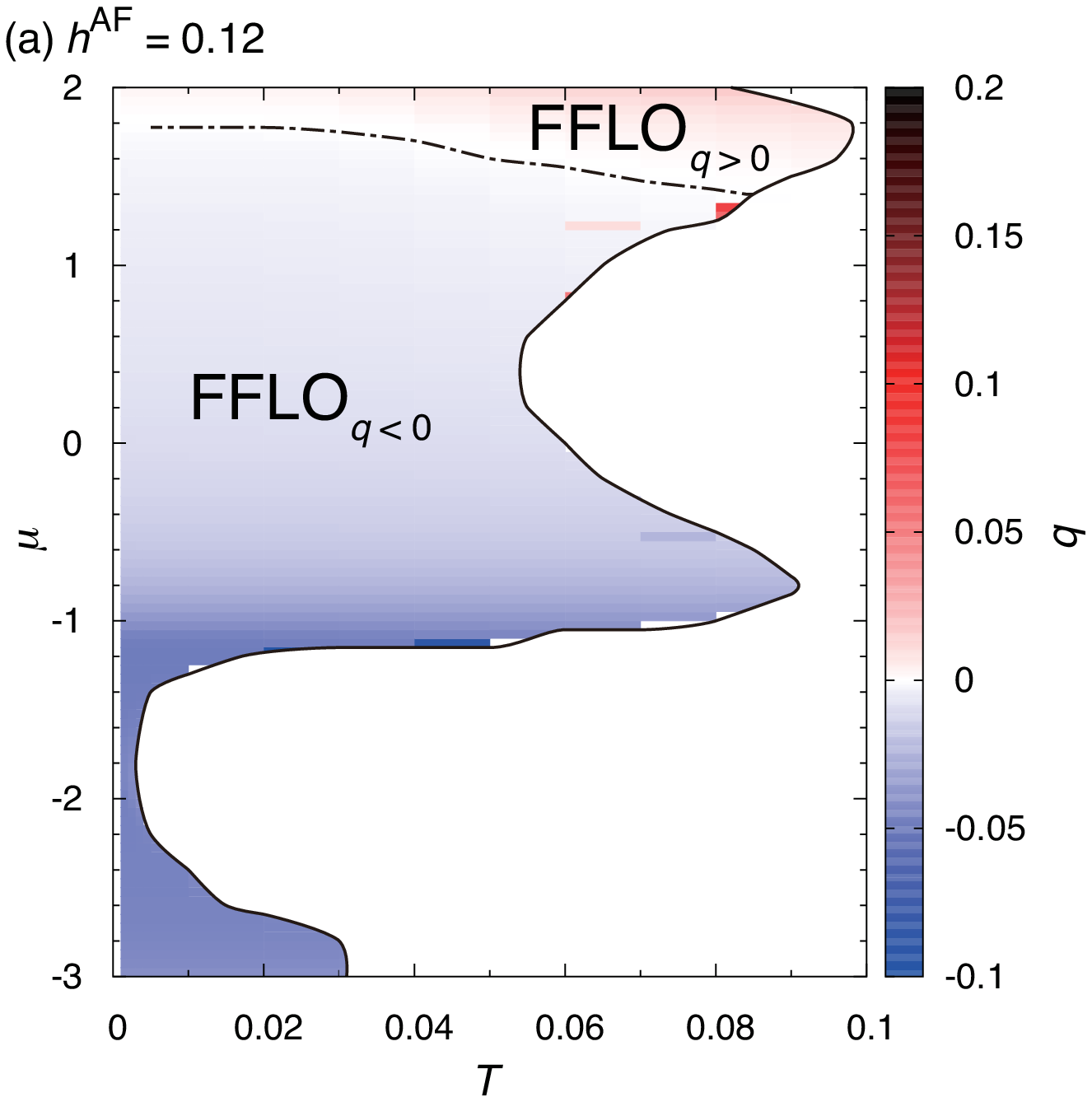}\\
 \includegraphics[width=75mm, clip]{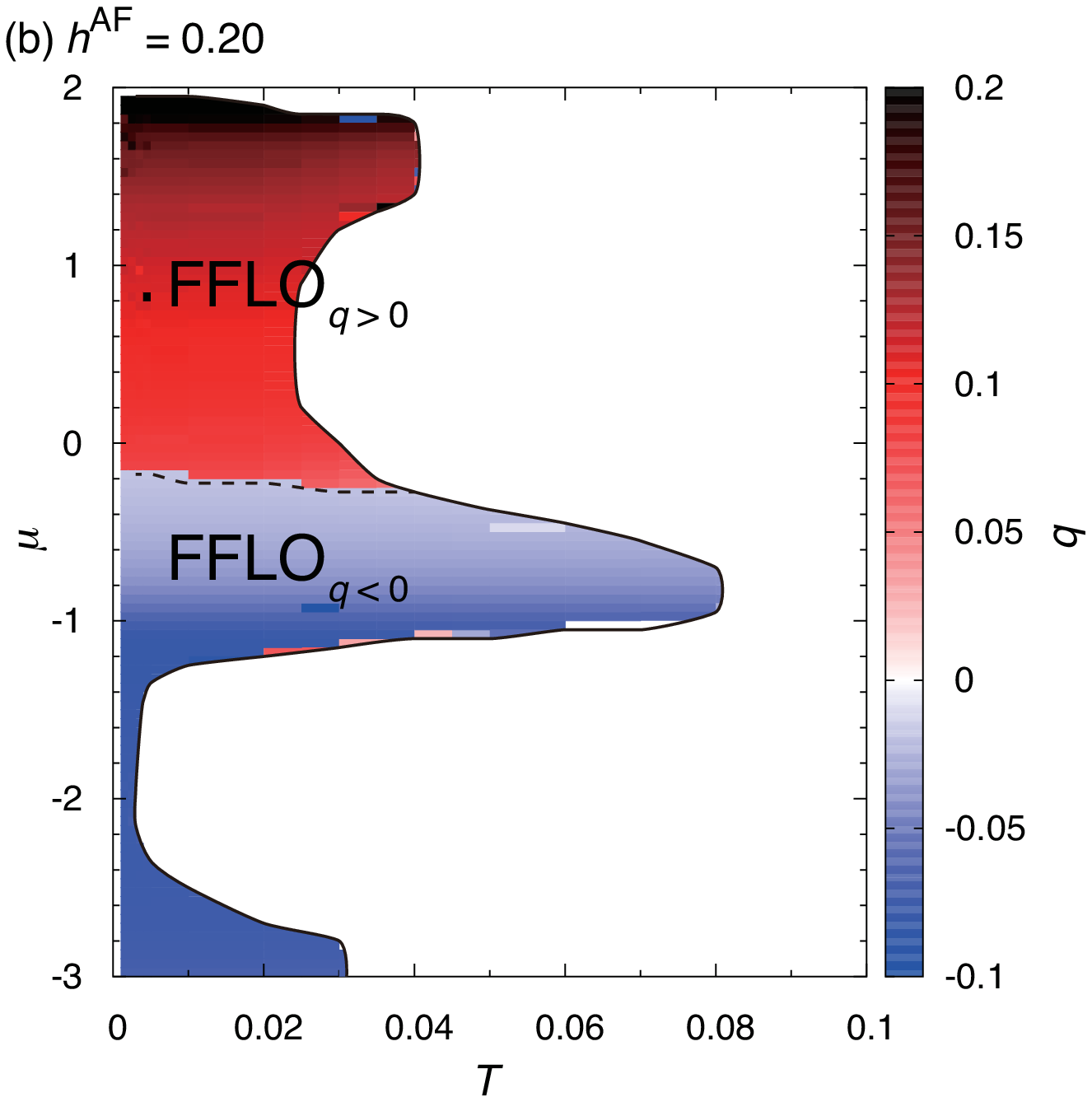}
 \caption{$T$-$\mu$ phase diagram in the magnetic quadrupole state for (a) $h^{\text{AF}} = 0.12$ and (b) $h^{\text{AF}} = 0.20$. The center-of-mass momentum of Cooper pairs $q$ is represented by color. The dashed line shows a first-order phase transition line, while the dash-dotted line shows a crossover line.}
 \label{fig:T-mu_phase}
\end{figure}

We address the $T$-$\mu$ phase diagram for two values of $h^{\text{AF}}$ in Fig.~\ref{fig:T-mu_phase}.
The Cooper pairs have finite center-of-mass momentum in the whole superconducting phase owing to the asymmetric band structure.
The asymmetry results from the symmetry of magnetic quadrupole state, and therefore, the FFLO state is stable irrespective of the parameters unless the ASOC vanishes.
When $\alpha = 0$, the band structure is symmetric and the BCS state is stable in a large parameter region.

Let us discuss the phase diagrams in details. 
We notice common features in Figs.~\ref{fig:T-mu_phase}(a) and \ref{fig:T-mu_phase}(b).
Critical temperature is rather higher for $\mu \gtrsim - 1$ than for $\mu \lesssim - 1$.
This is because the density of states (DOS) is large in the two-band region, $\mu \gtrsim - 1$.
The critical temperature is furthermore enhanced in the vicinity of the band edge ($\mu \simeq - 3, - 1, 2$) because of the large DOS.
Figure~\ref{fig:T-mu_phase} also reveals differences between the ``small quadrupole moment region'' ($h^{\text{AF}} = 0.12$) and the ``large quadrupole moment region'' ($h^{\text{AF}} = 0.20$).
In Fig.~\ref{fig:T-mu_phase}, we specify the FFLO state with $q > 0$ ($q < 0$) by ``FFLO$_{q > 0}$'' (``FFLO$_{q < 0}$'').
While the center-of-mass momentum $q$ continuously changes in the small quadrupole moment region, the FFLO$_{q < 0}$ state is separated from the FFLO$_{q > 0}$ state by the first-order phase transition line in the large quadrupole moment region [Fig.~\ref{fig:T-mu_phase}(b)].
The negative $q$ in the small $\mu$ region comes from the shift of the lower energy band to the negative momentum side.
The sum of the two Fermi momenta in the lower band is negative.
On the other hand, the upper band favors the FFLO$_{q > 0}$ state, and thus the FFLO$_{q < 0}$ state competes with the FFLO$_{q > 0}$ state in the two-band region.
As expected, the center-of-mass momentum increases with $\mu$ across the Lifshitz transition.
We show the $\mu$ and $T$ dependence of $q$ by color in Fig.~\ref{fig:T-mu_phase}.
We see the continuous change of $q$ in the small quadrupole moment region [Fig.~\ref{fig:T-mu_phase}(a)], while we observe a discontinuous jump at $\mu \simeq - 0.20$ in the large quadruple moment region [Fig.~\ref{fig:T-mu_phase}(b)].

\begin{figure*}[htbp]
 \centering
 \begin{minipage}[b]{60mm}
  \includegraphics[width=50mm, clip]{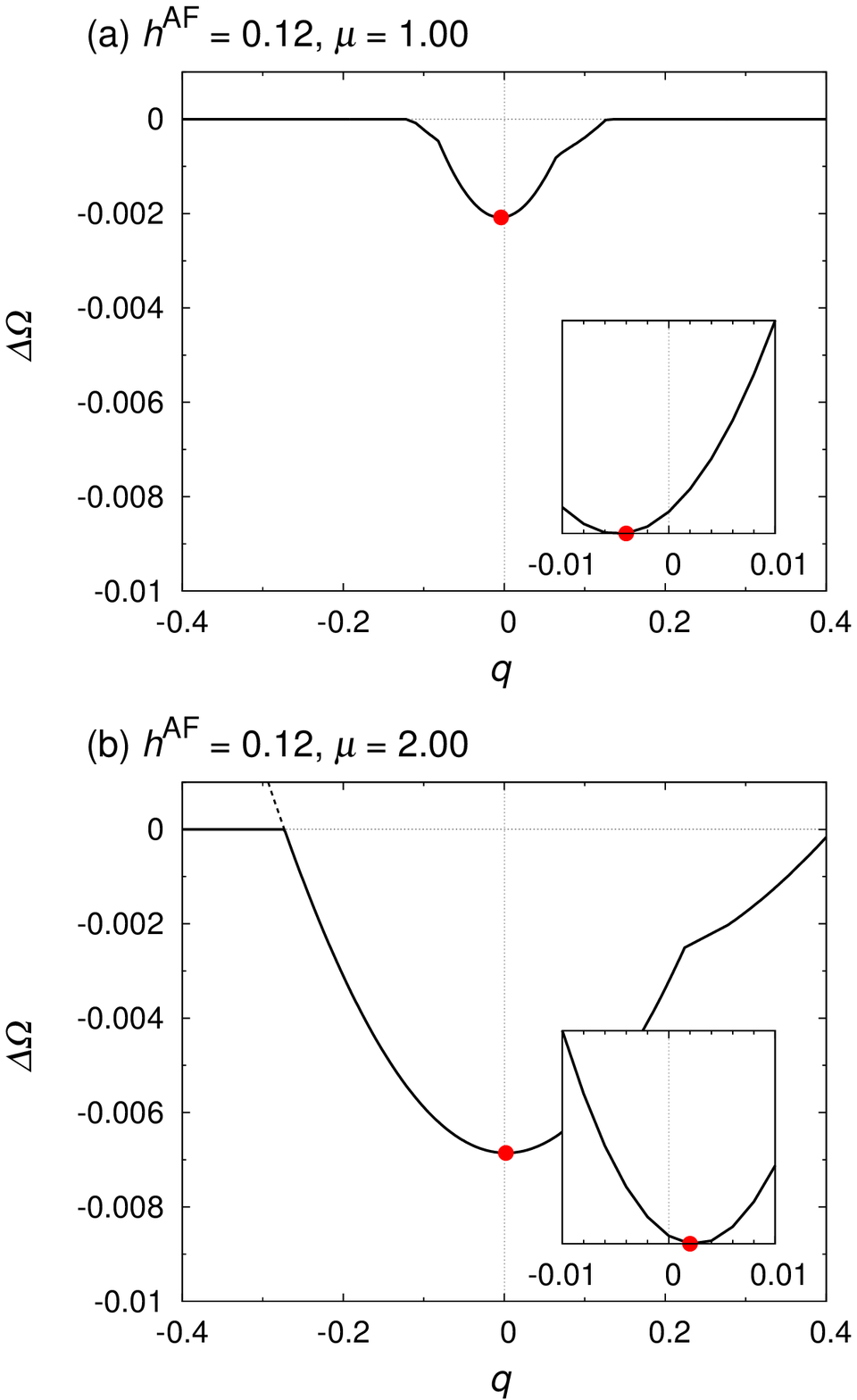}
 \end{minipage}%
 \begin{minipage}[b]{110mm}
  \includegraphics[width=100mm, clip]{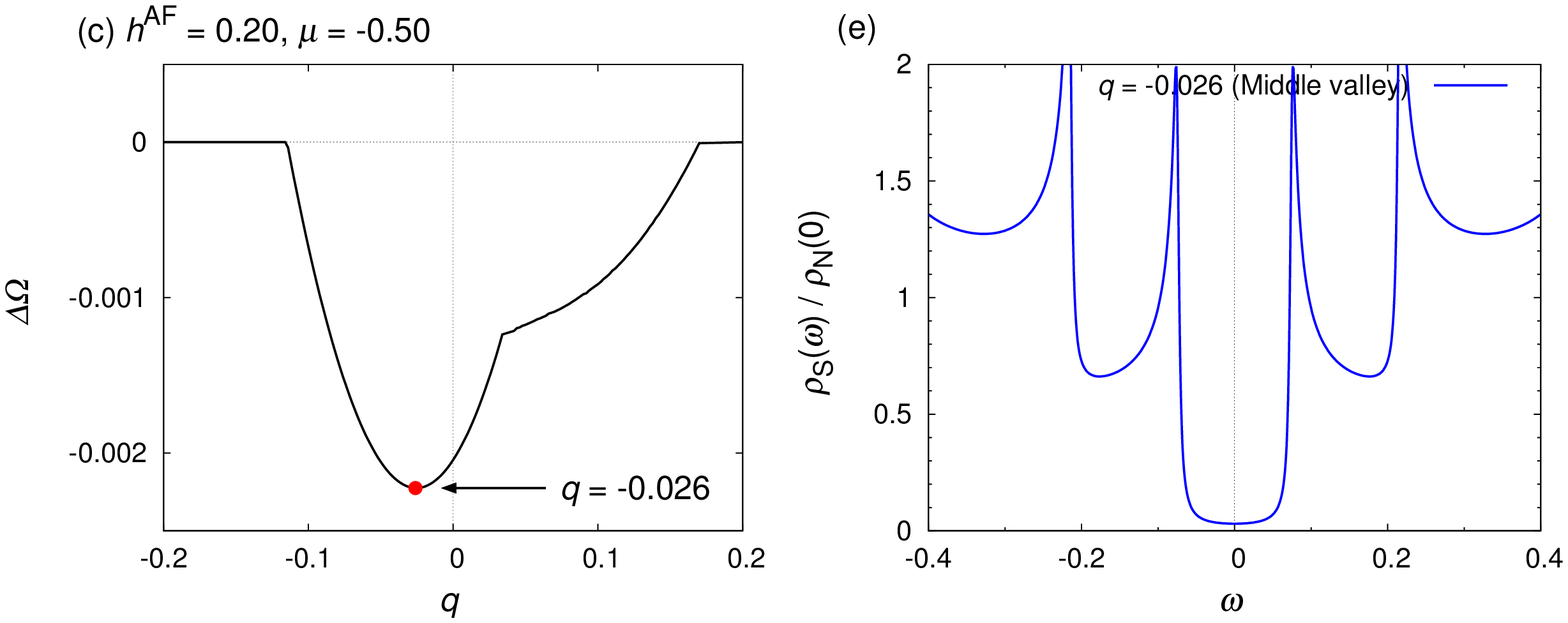}\\
  \includegraphics[width=100mm, clip]{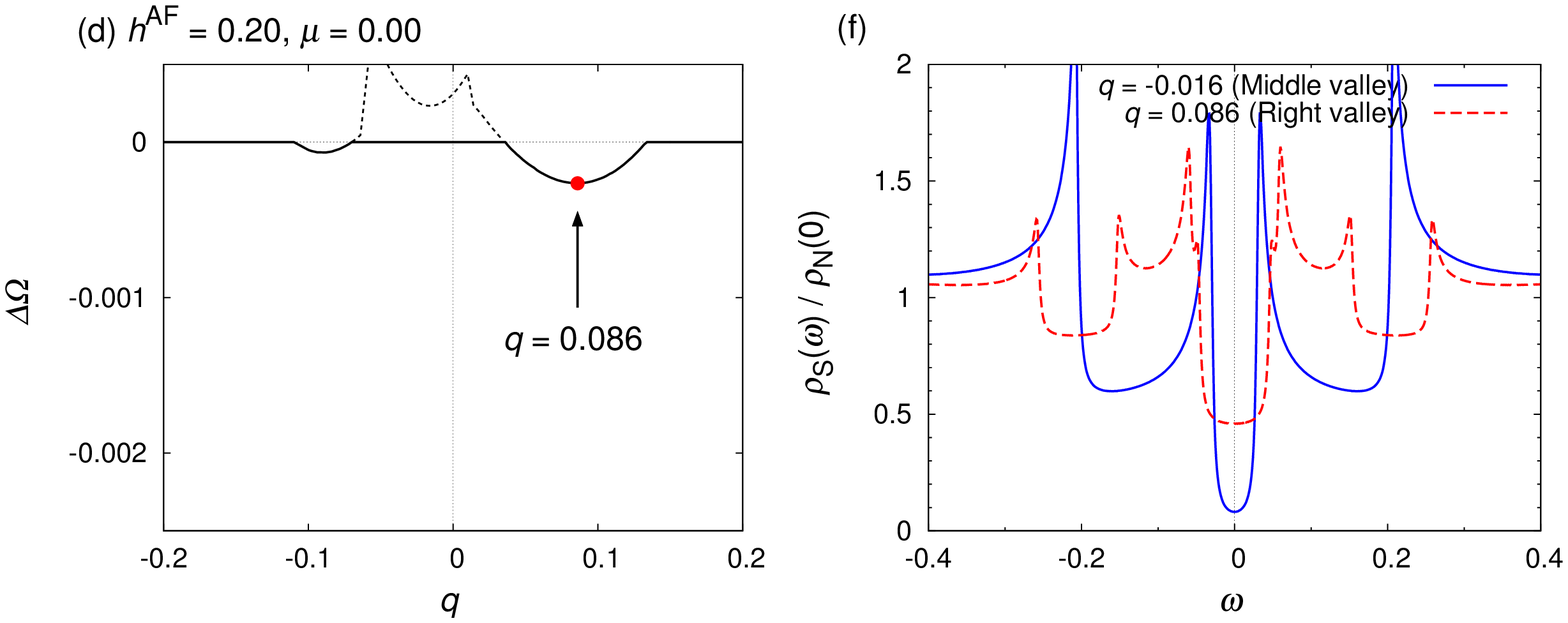}
 \end{minipage}
 \caption{(a) and (b) The $q$ dependence of the condensation energy $\Delta\Omega = \Omega_{\text{S}} - \Omega_{\text{N}}$ for a small quadrupole moment $h^{\text{AF}} = 0.12$ at $\mu = 1$ and $\mu = 2$, respectively. (c) and (d) $\Delta\Omega$ for a large quadrupole moment $h^{\text{AF}} = 0.20$ at $\mu = - 0.5$ and $\mu = 0$, respectively. The red points show the optimal $q$ which minimizes the condensation energy. (e) and (f) Superconducting DOS $\rho_{\text{S}}(\omega)$ normalized by the normal state DOS at the Fermi level $\rho_{\text{N}}(0)$ for the parameters in (c) and (d), respectively.}
 \label{fig:cond_energy}
\end{figure*}

\subsection{Condensation energy and DOS}
In order to elucidate what mainly determines the $q$ in the FFLO state, we look at the condensation energy $\Delta\Omega = \Omega_{\text{S}} - \Omega_{\text{N}}$, which is the difference of free energy between in the superconducting state and in the normal state.
The free energy in the normal state is obtained by just assuming $\Delta_a = \Delta_b = 0$. 

Figures~\ref{fig:cond_energy}(a) and \ref{fig:cond_energy}(b) show the condensation energy as a function of $q$ in the small quadrupole moment region.
Only one valley appears and its bottom moves to the positive-$q$ side with increasing $\mu$.
Thus, the optimal $q$ which minimizes the condensation energy continuously varies.

On the other hand, we find three valleys in the large quadrupole moment region.
In Fig.~\ref{fig:cond_energy}(d), the left and right valleys lead to a negative condensation energy, while the middle valley shows a positive condensation energy indicating a metastable state.
When we decrease the chemical potential to be $\mu = - 0.5$ [Fig.~\ref{fig:cond_energy}(c)], the condensation energy takes a minimum at $q = - 0.026$ which adiabatically changes to the bottom of the middle valley by increasing $\mu$.
This means that the FFLO$_{q < 0}$ state corresponds to the middle valley while the FFLO$_{q > 0}$ state corresponds to the right valley.
In other words, the center-of-mass momentum $q$ discontinuously changes because the valley structure appears in the free energy.
On the other hand, the valley structure is hidden and only the middle valley has a local minimum in the small quadrupole moment region.

Next we show the DOS of quasiparticles in order to clarify the superconducting states corresponding to the three valleys.
At both $\mu = - 0.5$ and $\mu = 0$, the DOS shows a superconducting gap near $\omega = 0$ in the ``middle-valley state'' [Figs.~\ref{fig:cond_energy}(e) and \ref{fig:cond_energy}(f)].
The narrower gap at $\mu = 0$ than at $\mu = - 0.5$ indicates that the ``middle-valley state'' is not likely to be stable at $\mu = 0$.
Indeed, Fig.~\ref{fig:cond_energy}(d) shows that the middle-valley state is metastable and the ``right-valley state'' is stable.
In contrast to the middle-valley state, approximately half of the DOS is residual at $\omega = 0$ in the right-valley state [Fig.~\ref{fig:cond_energy}(f)].
Thus, it is implied that although both energy bands contribute to the superconductivity in the middle-valley state, the upper (lower) band mainly causes the superconductivity in the right-valley (left-valley) state.
In other words, the lower band is weakly superconducting and gives rise to the large residual DOS in the right-valley state.
This view is consistent with the fact that the center-of-mass momentum $q$ in the right-valley state almost coincides with the sum of the Fermi momentum in the upper band.
Thus, the band-dependent FFLO state is stabilized by a large magnetic quadrupole moment.
Quasiparticles on the Fermi surface of the upper band form Cooper pairs, while the mismatch of $q$ and distorted lower band suppresses the superconducting gap in the lower band.
On the other hand, in the middle-valley state the superconductivity almost equivalently affects the two bands.
Then, $q$ is slightly negative because the distortion of the lower band is larger than that of the upper band.

Finally, we suggest an experimental test for the FFLO state.
The measurement of Josephson current in a FFLO SC/BCS SC junction may identify the single-$q$ FFLO state.
Since Josephson coupling vanishes in this junction due to the spatial modulation of the order parameter in the FFLO SC, the junction should carry a small Josephson current.
On the other hand, in an applied transverse uniform current in the BCS SC, a peak in the Josephson current may be found~\cite{Kim2016}.
The peak serves as an indicator of the FFLO state.

\section{Summary and discussion}
\label{sec:summary}
In this paper, we investigated the superconductivity in the magnetic multipole states.
In locally noncentrosymmetric systems with sublattice degree of freedom, not only the conventional magnetic dipole moment but also some odd-parity multipole moments may be polarized.
Ferroic multipole states with crystal momentum $q_{\text{M}} = 0$ were considered in the 1D zigzag chain as a minimal model.
Exotic superconducting states were elucidated as follows. 

The conventional BCS state is robust against the existence of ``antiferromagnetic moments'' in the unit cell which is regarded as a magnetic monopole.
Meanwhile in the dipole order the odd-parity spin-singlet PDW state is stabilized.
The situation in the latter corresponds to uranium-based heavy-fermion SCs UGe$_2$~\cite{Saxena2000}, URhGe~\cite{Aoki2001}, and UCoGe~\cite{Huy2007}.
It has been thought that the spin-triplet superconductivity occurs in these materials.
However, our result opens a new possibility that the ferromagnetic superconductivity in these materials is attributed to the PDW state.
From the theoretical point of view, the PDW state is identified to be a topological superconducting state when one of the bands is fully spin polarized.
We showed a nontrivial winding number in the class BDI, as well as nontrivial $\mathbb{Z}_2$ invariant in the class D.
The nontrivial topological numbers ensure the single Majorana end state.

Interestingly, the magnetic quadrupole order combined with the spin-orbit coupling makes the band structure asymmetric.
As a result of the asymmetric energy band, the FFLO state is stabilized without spin polarization.
This finding paves a new way for searches of the FFLO state~\cite{Matsuda_FFLO}.
Although previous studies researched SCs with a large Maki parameter~\cite{Bianchi2003, Kenzelmann2008, Uji2006, Lortz2007}, the external magnetic field applied to stabilize the FFLO state induces vortices which may obscure the FFLO state.
On the other hand, the FFLO state caused by the magnetic quadrupole order is free from the vortex.
Thus, a conclusive evidence for the FFLO state may be obtained by searching the superconductivity coexisting with the magnetic quadrupole order.

The band-dependent properties of the FFLO state were clarified as follows. 
When the magnetic quadrupole moment is small, the upper and lower energy bands are almost equally superconducting (if they cross the Fermi level).
Then, the center-of-mass momentum of the Cooper pair is small and continuously increases with chemical potential.
On the other hand, the center-of-mass momentum discontinuously changes in the large quadrupole moment region.
The origin of this first-order phase transition in the FFLO state is attributed to the band-dependent FFLO superconductivity.
While the two bands are almost equally superconducting at small chemical potentials, only the upper band mainly causes the superconductivity at large chemical potentials.
The two-band electronic structure is not an artifact of the 1D zigzag chain, but is a consequence of the nonsymmorphic crystal symmetry protecting the band degeneracy at the Brillouin zone boundary.
Therefore, the band-dependent FFLO superconductivity may be realized in various nonsymmorphic crystals hosting the magnetic quadrupole order.

\section*{Acknowledgments} 
The authors are grateful to T.~Arima, K.~Onozawa, M.~Sato, S.~Takamatsu, Y.~Nakamura, T.~Hitomi, and A.~Daido for fruitful discussions.
This work was supported by a ``J-Physics'' (15H05884) Grant-in-Aid for Scientific Research on Innovative Areas from MEXT of Japan, and by JSPS KAKENHI, Grants No. 24740230, No. 15K051634, No. 15H05745, and No. 16H00991.

\bibliography{paper}

\end{document}